\documentclass[reprint,aps,prl,twocolumn,superscriptaddress,footnoteinbib,nobalancelastpage,longbibliography]{revtex4-1}
\usepackage{graphicx}
\usepackage{{color}}
\usepackage{amsmath,amssymb,amsfonts}
\usepackage{braket}
\usepackage[colorlinks=True,citecolor=blue,linkcolor=red,urlcolor=blue]{hyperref}
\usepackage[mathscr]{euscript}
\newcommand{\up}{\uparrow}
\newcommand{\dn}{\downarrow}
\newcommand{\Up}{\Uparrow}
\newcommand{\Dn}{\Downarrow}

\newcommand{\sectionn}[1]{{\underline{{#1.}}}}
\newcommand{\sectionb}[1]{{\textit{{#1}}}}

\begin{document}
\title{On time crystallinity in dissipative Floquet systems}
\author{Achilleas Lazarides}
\affiliation{Max-Planck-Institut f\"ur Physik komplexer Systeme, N\"othnitzer Stra{\ss}e 38, 01187 Dresden, Germany}
\affiliation{Interdisciplinary Centre for Mathematical Modelling and Department of Mathematical Sciences, Loughborough University, Loughborough, Leicestershire LE11 3TU, United Kingdom}
\author{Sthitadhi Roy}
\affiliation{Physical and Theoretical Chemistry, Oxford University,
South Parks Road, Oxford OX1 3QZ, United Kingdom}
\affiliation{Rudolf Peierls Centre for Theoretical Physics, Clarendon Laboratory, Oxford University, Parks Road, Oxford OX1 3PU, United Kingdom}
\author{Francesco Piazza}
\author{Roderich Moessner} 
\affiliation{Max-Planck-Institut f\"ur Physik komplexer Systeme, N\"othnitzer Stra{\ss}e 38, 01187 Dresden, Germany}

\begin{abstract}
We investigate the conditions under which periodically driven quantum systems subject to dissipation exhibit a stable subharmonic response. Noting that coupling to a bath introduces not only cooling but also noise, we point out that a system subject to the latter for the entire cycle tends to lose coherence of the subharmonic oscillations, and thereby the long-time temporal symmetry breaking. We provide an example of a short-ranged two-dimensional system which does not suffer from this and therefore displays persistent subharmonic oscillations stabilised by the dissipation. We also show that  this is fundamentally different from the disordered DTC previously found in closed systems, both conceptually and in its phenomenology. The framework we develop here clarifies how fully connected models constitute a special case where subharmonic oscillations are stable in the thermodynamic limit.
\end{abstract}
\maketitle

\sectionn{Introduction\label{sec:intro}}
Understanding how statistical mechanics emerges in closed quantum many-body systems undergoing coherent dynamics with time-independent Hamiltonians has been one of the major themes of  physics research over the last few decades. More recently, attention has been focussed on closed systems with time-periodic (``Floquet'') Hamiltonians, where fundamentally novel out-of-equilibrium phases describable in macroscopic terms have been discovered; none more prominent than the $\pi$-spin glass also termed the discrete time crystal (DTC)~\cite{khemani2016phase,else2016floquet,moessner2017equilibration,Sacha2015,yao2017discrete,zhang2017observation,choi2017observation,pal2018rigidity,rovny2018observation,OSullivan2018}.

Generically, a major obstacle to working with Floquet systems is that they suffer heat death due to unbounded increase of entropy, approaching an infinite-temperature state at long times~\cite{dalessio2014long,lazarides2014equilibrium,ponte2015periodically}. This heating can be avoided by introducing disorder-induced localisation~\cite{lazarides2015fate,ponte2015many}, or by coupling the system to an external environment which drains energy from the system. The former approach, used in Ref.~\cite{khemani2016phase}, additionally endows the Floquet eigenstates with a discrete-symmetry broken spatial glassy order. Crucially, the eigenstates connected by the symmetry (and having the same spatial ordering pattern) are separated in quasienergy by $\pi/T$ with $T$ the driving period, leading to the subharmonic oscillation of an appropriate local observable.
It was shown in Ref.~\cite{lazarides2017fate} that an external Markovian environment, unless explicitly fine-tuned, destroys such a delicate coherence required for the subharmonic oscillations; the system is driven towards a mixture of various Floquet eigenstates all with uncorrelated patterns of the spatial glassy order.

\begin{figure}[t]
\includegraphics[width=1.\columnwidth]{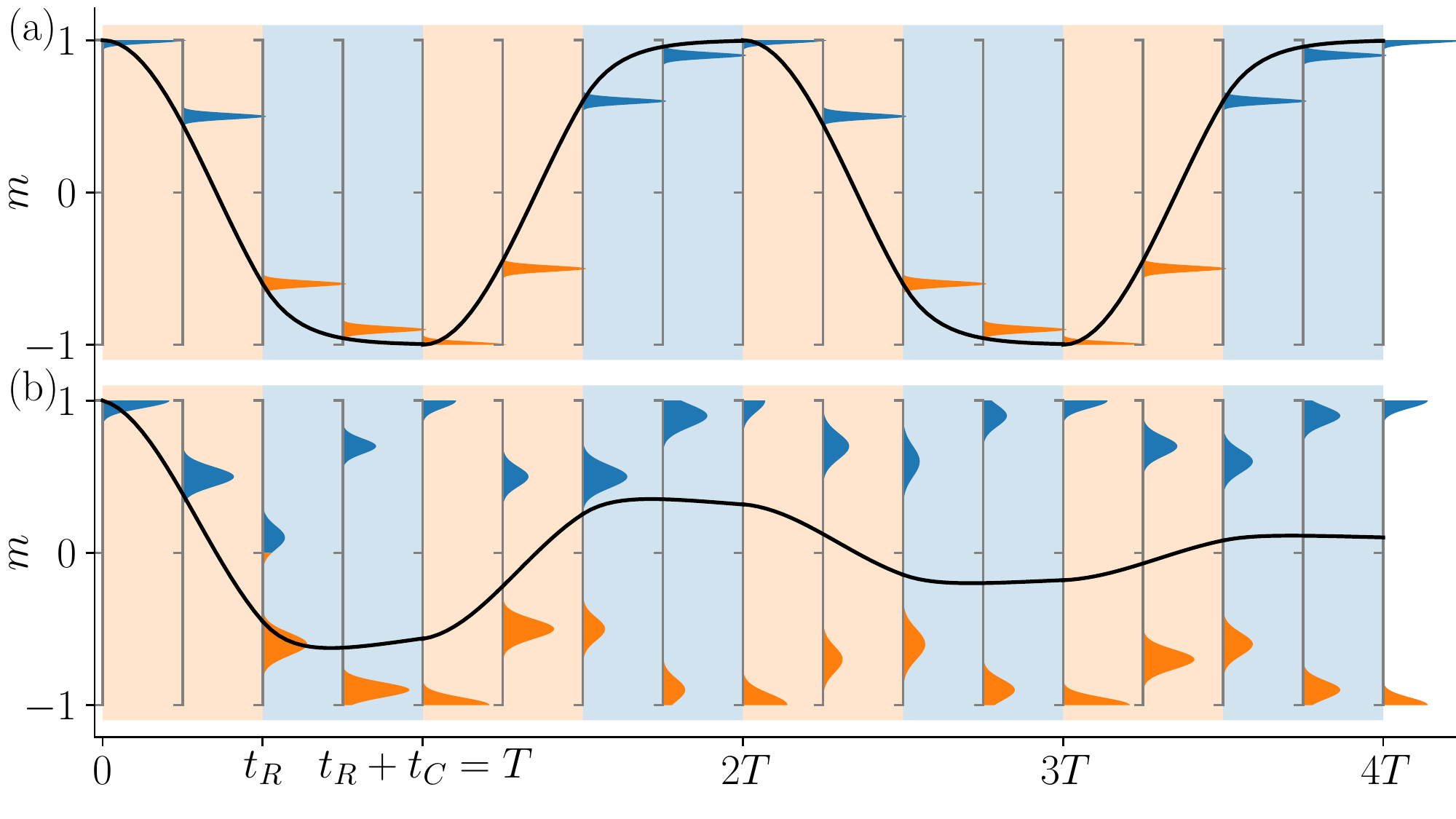}
\caption{
	{\bf{Schematic of how a dissipative DTC can (a) survive and (b) die:}} Evolution of the distribution and the mean (black line) magnetisation of a quantum spin system with time. Within a period (of $T$) a unitary process effecting global spin rotation and non-unitary cooling processes act together for $t_R$ whereas for $t_C$, the cooling processes act alone. (a) and (b) show an example where the magnetisation distribution does not broaden in time. Consequently in (a) the weight is always on the \emph{correct} magnetisation sector leading to the persistent subharmonic oscillations whereas in (b), the cooling processes split the distribution as such there is weight left behind in the \emph{wrong} sector which eventually kills the DTC.
	}
	\label{fig:broadening}
\end{figure}

Here we analyse general dissipative Floquet systems in a different setting where the ordered phases, time-crystalline or otherwise, are {\it stabilised} by the dissipation, and which would be entirely absent without it. The mechanism, manifestly different from that of the $\pi$-SG, involves a periodic rotation between two ``sectors'' of a Hilbert space followed by dissipative ``cooling'' to states distinguishable by a measure such as magnetisation.

This intuitively appealing picture ignores the possibility that the dissipation in addition to the desired cooling also generates noise and eventual loss of phase coherence in the oscillations. In terms of the density matrix of the system, the noise potentially translates to the probability distribution of the observable over individual realisations being broad, which we argue destroys the DTC. We show that while this is indeed the case for 1-$d$ spin chains with local interactions and dissipation, in 2-$d$ the dissipative processes can naturally lead to a narrow distribution and hence persistent time-crystalline order unlike in 1-$d$. We note that this broadening mechanism can be absent altogether in mean-field dynamics such as for fully-connected models~\cite{Russomanno2017,ging2018discrete,gambetta2019discrete,yao2018classical}.

\sectionb{How a dissipative DTC can exist$\ldots$}

First, let us describe the general arguments for the stability, or lack thereof, of DTCs to dissipation, before demonstrating them concretely using specific quantum spin-models. For a visual schematic, see Fig.~\ref{fig:broadening}.
\begin{itemize}
	\item The Hilbert space is divided into different (for concreteness: two) sectors, $\mathcal{H}_\pm$, which could be symmetry sectors or simply based on an empirical criterion based on the expectation value of an observable, and
	\item each of these sectors has a manifold of states $\{\ket{G_\pm}\}$ which posses \emph{quantum order} characterised by an observable, $\hat{M}$, which may for example be due to symmetry-broken order. Crucially,
	\item the two ordered manifolds, $\{\ket{G_\pm}\}$, cannot be connected to each other via local operators, and
	\item the expectation value $m$ of the order parameter, $\hat{M}$, is sufficiently narrowly distributed over the states within each of the manifolds that the two distributions for the two manifolds do not overlap (up to exponentially small corrections).
\end{itemize}
In Fig.~\ref{fig:broadening}, the two sectors are the positive- and negative-$m$ halves of the vertical axis. Within this setting consider a two-step Floquet protocol in the presence of dissipative processes such that,
\begin{itemize}
  \item In the first step (``rotation phase''), the system evolves under the simultaneous action of a Hermitian rotation operator $H_R$ and the dissipative terms dissipation. In the absence of dissipation the evolution over the step is unitary and given by $U_R(\theta)$, which maps neighbourhoods of the ground state manifold of one sector to states of the other sector and vice-versa. For a particular $\theta_\ast$, $U_R(\theta_\ast)$ maps states from $\{\ket{G_+}\}$  exactly onto states from $\{\ket{G_-}\}$ and vice versa.
\item In the second step (``cooling phase''), the system is governed by a Hamiltonian for which $\{\ket{G_\pm}\}$ are ground state manifolds as well as by the same dissipation processes as in the first step.
\end{itemize}
The dissipative processes during both steps cool the system down so that under their sole influence all states in the $\mathcal{H}_+$ sector would be driven to $\{\ket{G_+}\}$ and likewise for $\mathcal{H}_-$. Under the combined action of the unitary and dissipative terms, they take the system towards the $\ket{G_{\pm}}$ of the sector it instantaneously finds itself in.

A quantum system driven with such a dissipative Floquet protocol, initialised in either of the ground state manifolds, shows a time crystalline response trivially if $\theta=\theta_\ast$, as the expectation value of $\hat{M}$ oscillates stroboscopically between that in $\{\ket{G_+}\}$ and $\{\ket{G_-}\}$ with a period twice that of the Floquet drive, \textit{provided the dissipation is inactive in the rotation cycle}. A certain robustness of the temporal order to deviations of $\theta$ from $\theta_\ast$ is expected to be induced via the dissipation: if the unitary rotation does not take states from the ground state manifold of one sector (say $\{\ket{G_+}\}$) entirely to that of the other sector ($\{\ket{G_-}\}$), but admixes nearby excited states in the other sector, the cooling step of the drive can push the weight back onto the $\{\ket{G_-}\}$ manifold. Simply put, cooling kills off the excitations left behind by the imperfect rotation, stabilising a DTC.

\sectionb{$\ldots$ and how it can die:}
In a broad sense, the dissipative processes have three effects:
\begin{itemize}
	\item \emph{Hindering rotation during rotation phase} Recall that $\{\ket{G_+}\}$ and $\{\ket{G_-}\}$ are not connected via local operators. $U$ then naturally has the form of a global rotation of the degrees of freedom. If the rotation phase is not instantaneous, the state goes through excited states at intermediate times. However, the dissipation \emph{cools} the system, opposing this creation of excitations hence making the rotation process less effective. Therefore the overall rotation with dissipation is less than without, trapping (part of) the weight in the \emph{wrong} sector. This is unfavourable to the presence of a stable DTC.
	\item \emph{Correcting error caused by imperfect rotation during cooling phase} Imperfect rotation potentially leaves the state in the correct sector, but not in the $\ket{G_\pm}$ manifolds; dissipation corrects this, favouring the DTC. 
	\item \emph{Broadening the distribution during both phases}: The rotation and the cooling acting in conjunction can, for short times, increase the width of the distribution of the state's overlaps with the excited states such that the resulting state is spread over both the sectors. In the following cooling cycle of the Floquet drive, the weights in each sector can get pushed to their respective ground state manifolds, resulting in a finite weight in the $\emph{wrong}$ sector (see later discussion and Fig.~\ref{fig:broadening}). This is generally fatal to the DTC.
  \end{itemize}

Of the three, the third (broadening) invariably causes the DTC signal to decay eventually. In its absence, when the probability distribution of the observable remains sharp over time, a stable DTC phase is possible with the first two mechanisms determining the parameter regime of the stability. Let us also note that while dissipation is favourable for the temporal order in the cooling cycle, it is detrimental in the rotation cycle, and 
it is {\textit{a priori}} not obvious whether increasing the strength of the dissipation from some finite value favours or disfavours the temporal order. 

In what follows, we introduce an explicit microscopic model and give three examples of dissipative processes. First we show that dissipation that cleanly separates the two sectors but broadens the magnetization distribution leads to a decay of the oscillation. We then introduce spatially local dissipation processes and show that: 1) in 1-$d$ they fail to separate the two sectors, cause broadening, and lead to a decaying oscillation. In 2-$d$, they may cleanly separate the sectors and in addition do not result in broadening, so that in this case a stable DTC appears.

\sectionn{Quantum spin systems}
\label{sec:isinggeneral}
To analyse the above ideas in a concrete setting, we consider a system of spins-1/2, first in 1-$d$. Using a basis constituted by the products states of $\sigma^z$ (which we henceforth denote as $\{\ket{\alpha}\}$), the two sectors $\mathcal{H}_\pm$ can be taken to be the set of product states, $\{\ket{\alpha_\pm}\}$ which satisfy $\braket{\alpha_\pm|\hat{M}|\alpha_\pm}\gtrless 0$ respectively with $\hat{M}=\sum_\ell \sigma^z_\ell$~\footnote{Modulo the ambiguity for the product states with zero magnetisation; we choose to put half of them in the first sector and half of them in the other. The choice has no bearing on the subsequent dynamics.}. 
This is a natural choice for a system described by a ferromagnetic Ising Hamiltonian
\begin{equation}
\hat{H}_\mathrm{TFIM} = -\sum_\ell \sigma^z_\ell\sigma^z_{\ell+1} +g\sum_{\ell}\sigma^x_\ell,
\label{eq:isingham}
\end{equation}
as in the limit of $g\to 0$, $\{\ket{\alpha}\}$ is a possible set of eigenstates. Moreover, for $g\neq0$, $\ket{G_\pm}$ are adiabatically connected to the $\ket{\Up}$ (all-up) and $\ket{\Dn}$ (all-down) states as long as the Hamiltonian is in the ferromagnetic phase, $\vert g\vert<1$.
Note that, defining the two sectors and the corresponding ground state manifolds in the fashion we do, also allows us to label the basis states and the sectors with the magnetisation density $m=\braket{\hat{M}}/N$ ($N$ being the system size).

The unitary operator $U_R$ which in the thermodynamic limit maps states $\{\ket{G_+}\}\leftrightarrow\{\ket{\alpha_-}\}$ is given by $U_R(\theta)=\exp\left[-i\theta\sum_\ell\sigma^x_\ell\right]$ with $\theta\in(\pi/4,\pi/2]$ and is produced by the action of the Hamiltonian $H_R=\frac{\theta}{t_R}\sum_j\sigma^x_j$ over time $t_R$. It easily follows that for $\theta_\ast=\pi/2$, $U_R(\theta_\ast)$ precisely maps the all-up state to the all-down state. In fact, since the ground state of the system breaks the $\mathbb{Z}_2$ symmetry of the Hamiltonian spontaneously, $U_R(\theta_\ast)$ connects the two ground states exactly throughout the ferromagnetic phase.
As anticipated, the rotation $U_R(\theta)$ is manifestly a non-local operation. In the thermodynamic limit, a product state with definite magnetization $m$ is mapped to a (in the $\sigma^z$-basis, non-product) state with definite magnetization $m\cos(2\theta)$, so that the rotation operation does not result in broadening.

Depending on the dissipative processes involved, this model can show decaying (bottom of Fig.~\ref{fig:broadening}) or persistent (top) subharmonic oscillations depending on whether broadening occurs or not. In the following, we use three explicit Markovian dissipative processes to show how the absence (presence) of broadening due to them favours (disfavours) the persistence of the temporal order.

\sectionn{Lindblad dynamics}
\label{sec:lindblad}
Focussing on Markovian dissipative processes, the equation of motion for the density matrix of the system is governed by the Lindblad equation
\begin{equation}
\partial_t\rho = -i[\hat{H}(t),\rho]+\sum_i\left(\hat{L}_i\rho \hat{L}_i^\dagger-\frac{1}{2}\{\hat{L}_i^\dagger \hat{L}_i,\rho\}\right)
\end{equation}
where $H(t)$ is the time-dependent (in our case, time-periodic) Hamiltonian and $\{L_i\}$ is the set of time-independent quantum jump operators which arise due to the coupling to the dissipative environment. Our binary Floquet protocol with  period $T=t_C+t_R$ is
\begin{equation}
\hat{H}(t) = \begin{cases}
				\frac{\theta}{t_R}\sum_\ell\sigma^x_\ell;\, 0\le t<t_R\\
                \hat{H}_\mathrm{TFIM};\, t_R\le t<t_R+t_C \ .
             \end{cases}
\label{eq:timeperiodicham}
\end{equation}
We  focus mostly on the $g\to 0$ limit of $\hat{H}_\mathrm{TFIM}$.

\sectionn{Direct jump operators}
To demonstrate the deleterious effects of broadening we begin by considering a set of jump operators $\{L_\alpha\}$ 
\begin{equation}
L_\alpha = \sqrt{\gamma}\left[\Theta(m_\alpha)\ket{\Up}\bra{\alpha} + \Theta(-m_\alpha)\ket{\Dn}\bra{\alpha}\right],
\label{eq:dumpdef}
\end{equation}
where $m_\alpha$ denotes the magnetisation of the product state $\ket{\alpha}$.
These jump operators take the weight from any product state and transfer it \emph{directly} to the ground state of the corresponding sector as well as causing  exponential decay of offdiagonal elements of the density matrix in the product state basis. They therefore provide very efficient cooling. However, 
they lead to broadening of the distribution leading to the decay of the oscillatory signal. To show this explicitly, we study the time-dependent magnetisation of the system starting from the $\ket{\Up}$ state, making use of a simplification due to translational invariance to access very large system sizes \cite{supp}.

\begin{figure}
	\includegraphics[width=\columnwidth]{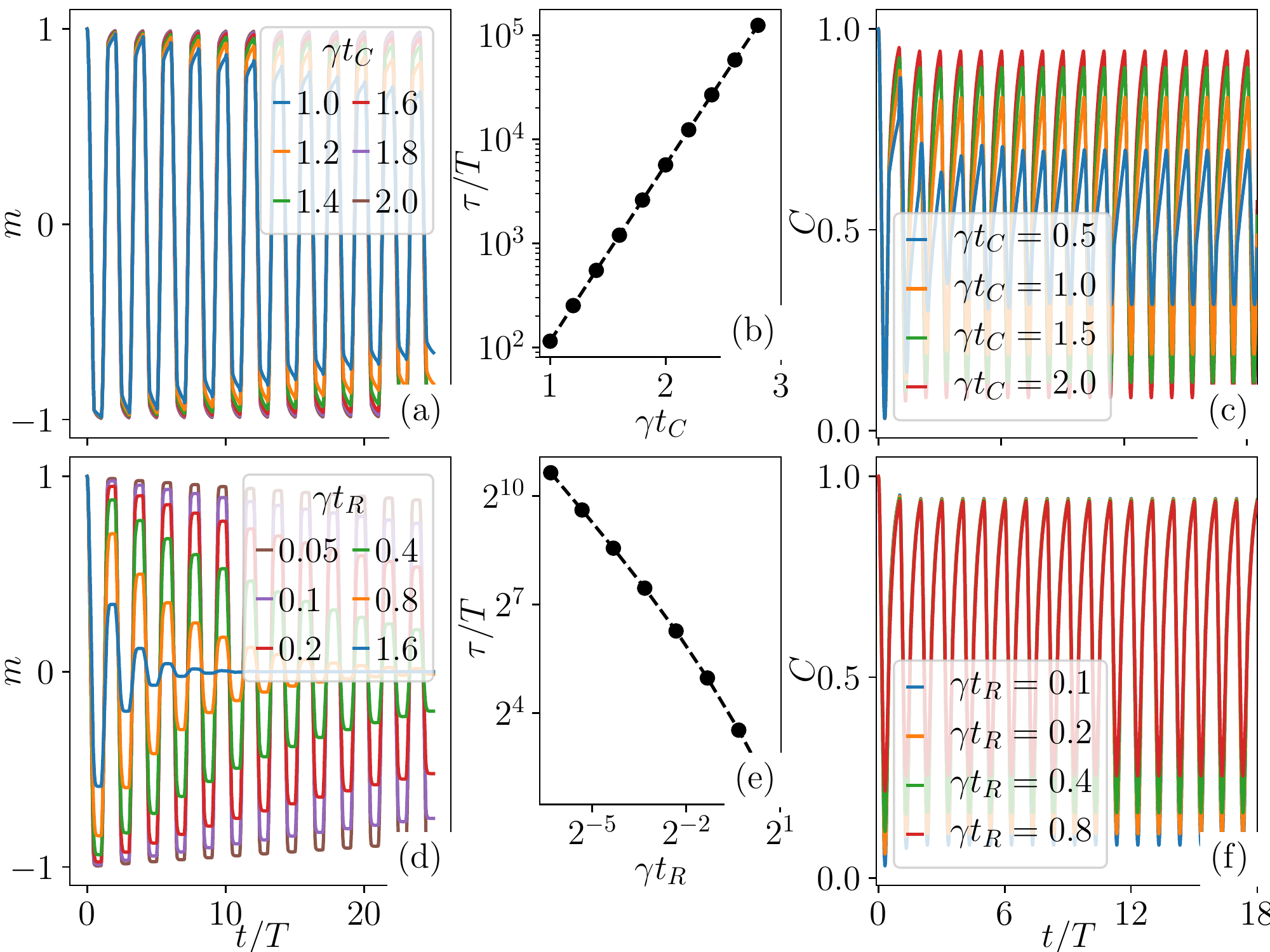}
	\caption{{\bf Direct jump operators}: The first column shows the {\bf instability} of the subharmonic oscillations of the magnetisation density, $m$, for the jump operators $\eqref{eq:dumpdiff}$ for (a) $\gamma t_R=0.1$ and different values of $\gamma t_C$ and (d) $\gamma t_C=10$ and different values of $\gamma t_R$. The respective lifetimes, $\tau$, are shown in the second column, revealing an exponentially increasing and at least polynomially decreasing lifetime with $\gamma t_C$ and $\gamma t_R$ respectively. The data is for $N=51$, and $\theta=0.45\pi$ and $\theta=\pi/2$ for the top and bottom rows respectively.
	{\bf Stability} of the persistent oscillations of the correlation function $C$, Eq.~\eqref{eq:correlation} is shown in the third column. For a fixed value of $\gamma t_R$ ($=0.1$ in (c)), the amplitude of the oscillations increases with $\gamma t_C$ as dissipation corrects the error induced by imperfect rotationms. By contrast, for a fixed $\gamma t_C$ ($=2$ in (f)), the amplitude decreases with $\gamma t_R$ as dissipation hinders the rotation. Here $N=51$ and $\theta=0.4\pi$.}
	\label{fig:dumpopsall}
\end{figure}


Fixing one of $\gamma t_R$ and $\gamma t_C$ and varying the other leads to the results shown in the two leftmost columns of Fig.~\ref{fig:dumpopsall}: for either finite $\gamma t_C$, finite $\gamma t_R$, or both finite, the time-crystalline response of $m$ decays exponentially with $t$ while the \emph{lifetime} of the subharmonic oscillations grows exponentially with $\gamma t_C$. On the other hand it decays at least polynomially with $\gamma t_R$, so overall stronger dissipation has opposite effects during each part of the driving. Nevertheless broadening of the distribution means the oscillations always decay via the mechanism shown in the lower panel of Fig.~\ref{fig:broadening}. 

The magnetisation vanishes with time due to the state strobscopically being in a mixture of both ground state manifolds, with opposite magnetisations.

One then expects that an observable finite and equal in both the ground states will remain finite in a statistical mixture of the two, such as is obtained at long times in the present case. Such an observable is the correlator
\begin{equation}
	C(t) = \frac{1}{N^2}\sum_{\ell\neq r}\mathrm{Tr}\left[\rho(t)\sigma^z_\ell\sigma^z_r\right].
	\label{eq:correlation}
\end{equation}
The persistent oscillations of $C(t)$ are shown in the rightmost column of Fig.~\ref{fig:dumpopsall}. This is a fundamental difference between this dissipative Floquet phase and the $\pi$-spin glass, where persistent oscillations of $C$ imply those of an initially finite $m$. The amplitude of the oscillations decreases with decreasing $\gamma t_C$ because the signal is strongest in the two fixed point states $\ket{\Up/\Dn}$ and $\gamma t_C$ controls how well the system is cooled into the two ground states. In this sense, this order is also 
$\emph{induced}$ by dissipation. Increasing $\gamma t_R$ causes $C$ to remain close to unity (its value in the ground states) and resist rotation, consistent with the earlier general arguments.

The direct jump operators demonstrate that the broadening of the distribution in magnetisation is fatal to the time-crystalline order, \emph{even when the dissipation cleanly separates the two sectors}.  

\sectionn{Domain-wall annihilating jump operators} We now introduce a set of jump operators which avoid broadening in a natural way. These operators cause dynamics that only move domain walls (DWs): a free standing DW can move but not disappear. However two DWs can move into each other and annihilate. Such dynamics are fundamentally different in the 1- and 2-dimensional cases.

Denoting the neighbours of a site $\ell$ by $\{r_\ell\}$, to each product state $\ket{\alpha}$ and site $\ell$ there corresponds a jump operator
\begin{equation}
  L_{\alpha,{\ell}} =\sqrt{\gamma}\left(\prod_{i\neq \ell}\mathcal{P}_{i,s^\alpha_i}\right)\times
  \begin{cases}
    \sigma^x_\ell,\quad M_{\{r_\ell\},\alpha}=0\\
    \sigma^\pm_\ell,\quad \mathrm{sgn} \left(M_{\{r_\ell\},\alpha}\right)=\pm 1
  \end{cases} 
\label{eq:localops}
\end{equation}
where $M_{\{r_\ell\},\alpha}$ is the net magnetisation of the spins in $\{r_\ell\}$ and $\mathcal{P}_{i,s_i^\alpha}$ is a projector onto the spin at site $i$ in spin-state $s_i^\alpha=\up/\dn$. The corresponding Lindblad dynamics along the diagonal is governed by a Pauli master equation
 \begin{equation}
\partial_t \rho_{\alpha\alpha} = \sum_\beta \gamma_{\alpha \beta}\rho_{\beta\beta}-\left(\sum_{\beta}\gamma_{\beta\alpha}\right)\rho_{\alpha\alpha}
\label{eq:masterequation}
\end{equation}
with the $\gamma$ determined according to the rules above while the off-diagonals decay exponentially. 

\begin{figure}
	\includegraphics[width=\columnwidth]{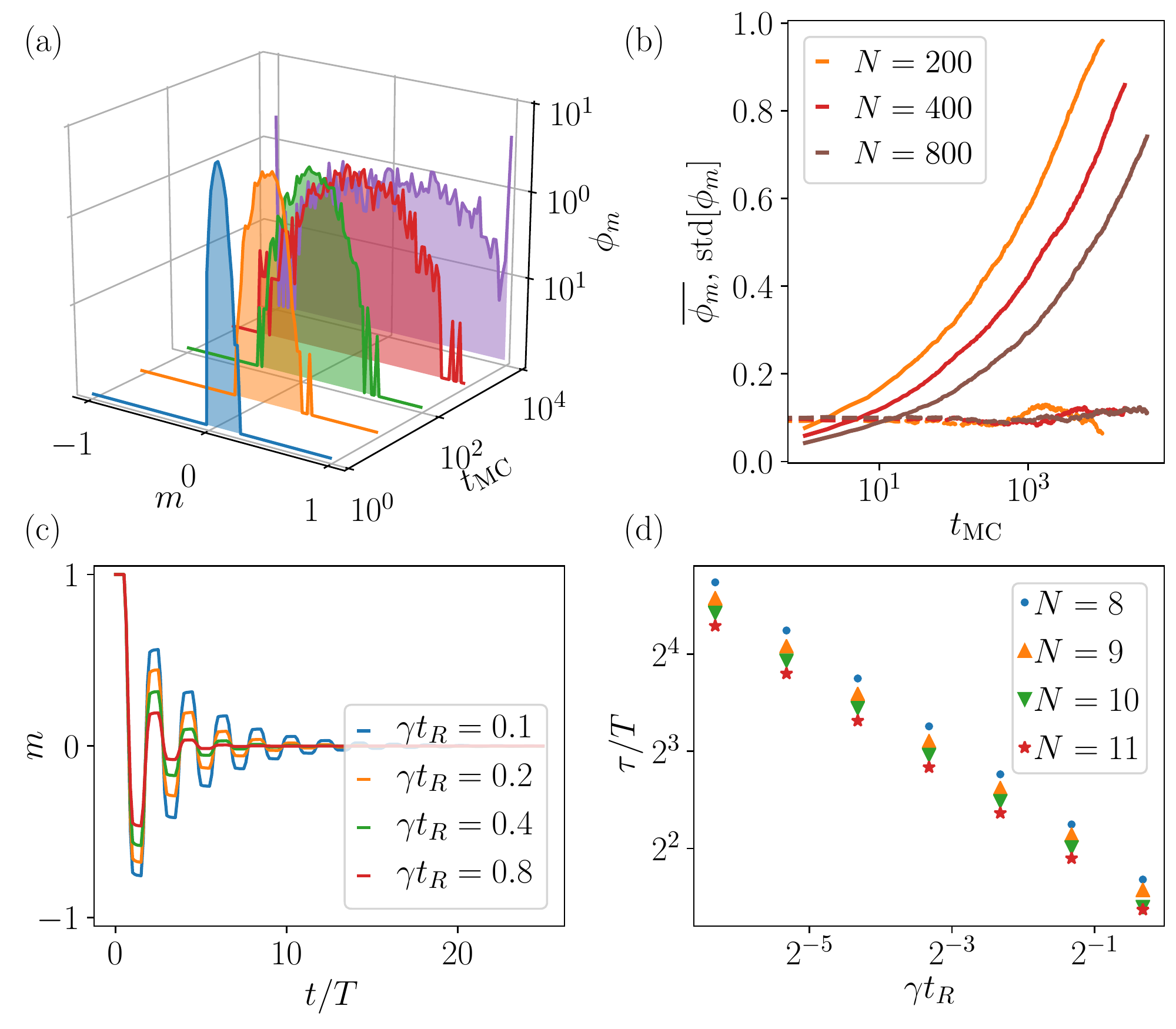}
	\caption{{\bf Instability of the DTC to domain wall annihilating operators in 1D}: (a) Results obtained from classical Monte Carlo in one dimension show that the distribution of the state in magnetisation for the jump operators in Eq.~\eqref{eq:localops} broadens with time. (b) The mean magnetisation (dashed lines) stays constant over Monte Carlo times whereas the standard deviation (solid lines) grows, indicating broadening of the distribution. (c)-(d) Numerically solving the Lindblad equation with the time-periodic Hamiltonian, \eqref{eq:timeperiodicham}, and the jump operators \eqref{eq:localops} shows an exponential decay of the time-crystalline order with polynomially decreasing lifetime with $\gamma t_R$. For the numerical solutions, $\theta=\pi/2$ and $\gamma t_C=100$.}
	\label{fig:local1d}
\end{figure}

In 1-$d$, the dynamics along the diagonal amounts to the DWs executing a random walk, {\it i.e.}~diffusing. The
probability distribution of magnetisation starting from a sharp value $m$ broadens at short times 
(Fig.~\ref{fig:local1d} top left), and at long times 
becomes bimodal with two peaks at $\pm 1$ of height such that $-p(-1)+p(1)=m$ (Fig.~\ref{fig:local1d} top right), as found  by solving Eq.~\eqref{eq:masterequation} using a classical kinetic Monte Carlo approach. The resulting destruction of the DTC
is shown in the bottom two panels.

\begin{figure}
\includegraphics[width=\columnwidth]{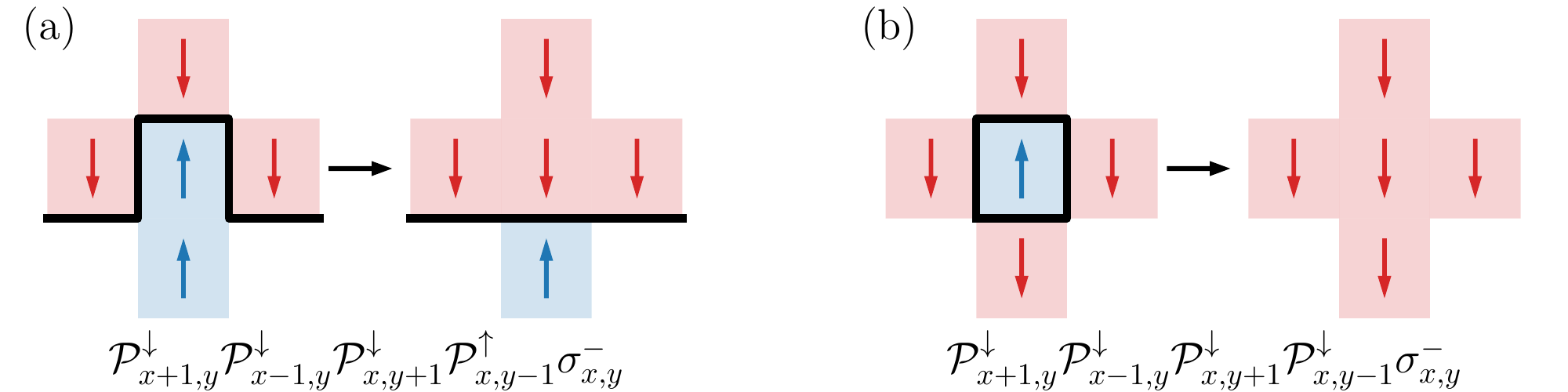}
\caption{{\bf Examples of the local jump operators leading to a persistent DTC in two dimensions:} In general and for individual product states, these operators tend to either decrease the length of domain walls, such as the transition shown on the left, or cause minority regions to vanish, as on the right.}
\label{fig:localops-2d_lattice}
\end{figure}

\begin{figure}
\includegraphics[width=\columnwidth]{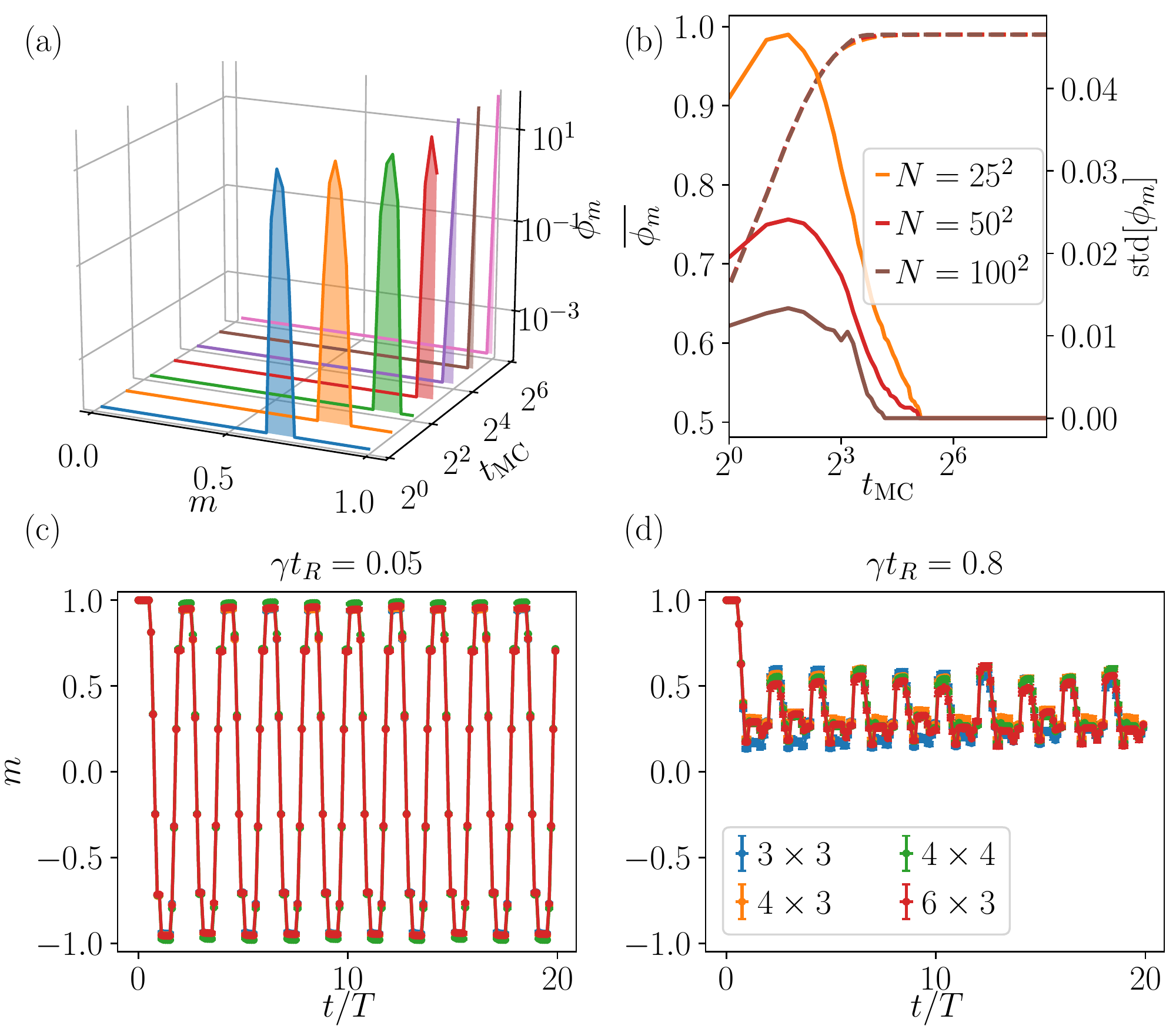}
\caption{{\bf Stability of the DTC to domain wall annihilating operators in 2D}: (a)Results obtained from classical Monte Carlo in two dimensions show that the distribution of the state in magnetisation for the jump operators in Eq.~\eqref{eq:localops} stays sharp. (b) The mean magnetisation (dashed lines) saturates to unity whereas the standard deviation of the system (solid lines) systematically goes down with system size. (c)-(d) Numerically solving the Lindblad equation for square lattices (of size $N_x\times N_y$) shows a persistent time-crystalline response of the magnetisation of for low $\gamma t_R$ whereas an oscillating ferromagnet at high $\gamma t_R$ For the numerical solutions, $\theta=\pi/2$, $\gamma t_C=100$, and $\gamma^\prime=\gamma/10$.}
\label{fig:local2d}
\end{figure}

For $d\geq2$, this type of domain wall dynamics eventually eliminates the minority phase by effectively causing a line (or surface) tension, tending to minimise the area of the interface between two non-conserved phases (see Fig.~\ref{fig:localops-2d_lattice} for two examples of allowed transitions and Sup.~Mat.~\cite{supp} for a demonstration of how the dynamics minimises the interface length) so that the dissipation cleanly separates the two sectors. In general, local dissipative processes lowering the energy of (ferromagnetic) Ising-type hamiltonians in $d\geq 2$ will behave in a qualitatively similar way.  Less obviously, the dissipative dynamics does not broaden the magnetization distribution starting from a state sharp in magnetization, Fig.~\ref{fig:local2d}. It then follows that a DTC phase may be stable; this is supported by the lower panels of Fig.~\ref{fig:local2d} where a rapid rotation is shown to result in persistent subharmonic oscillations while a slow rotation in a ferromagnetic phase in which the magnetization never changes sign.

In order to show this explicitly on finite-sized systems a numerical solution of the full Lindblad equation is desirable. However note that Eq.\eqref{eq:localops} dictates that the number of jump operators is exponentially large in $N$, making a numerical solution all but impossible. We surmount this by noting that one can have another set of jump operators which lead to identical (to those of Eq.\eqref{eq:localops}) dynamics for the diagonal elements of the density matrix but whose number grows polynomially with $N$. 

Each `majority rule' operator, for a given site, 
consists of a product of projectors onto the site's
neigbouring spins (the set is over all possible configurations of
$\mathcal{P}^{\up/\dn}$ for the neighbours) multiplied by the spin
raising (lowering) operator for the site if the projector
configuration has fewer $\mathcal{P}^{\dn(\up)}$ compared to
$\mathcal{P}^{\up(\dn)}$, see
e.g.\ Fig.~\ref{fig:localops-2d_lattice}. The resulting dynamics is
displayed in Fig.~\ref{fig:local2d}

.

\sectionn{Conclusions and outlook}
We have discussed general mechanisms leading to dissipative (de)stabilisation of DTCs in  {\it  disorder-free} dissipative Floquet systems. Our example of a dissipation-stabilised DTC is completely distinct, relying on a fundamentally different mechanism, from the $\pi$-spin glass introduced in~\cite{khemani2016phase}, which in turn is unstable to dissipation~\cite{lazarides2017fate}. We also uncover other, non-DTC but still dissipation-induced phases.

Phenomenologically, a crucial difference between the dissipative Floquet system and the $\pi$-SG~\cite{khemani2016phase} is that in the former the oscillations in the magnetisation can decay even though its correlation function, $C(t)$, can synchronise and oscillate persistently, while in the latter one implies the other.

While the mechanism of obtaining period-doubling from periodic switching between distinct sectors of Hilbert space is intuitively transparent, our analysis of its failure modes we believe also sheds light on recent work finding stable subharmonic oscillations~\cite{Russomanno2017,ging2018discrete,gambetta2019discrete}. In these works the system Hamiltonian is fully connected. This typically leads to a stochastic description in which the noise vanishes with diverging system size, so that the master equation for the density matrix results in no broadening over the time
evolution~\cite{gelhausen2018dissipative,benatti2018quantum}.  

We believe that generally, treatments for short-range models based on approximate mean-field and other analyses involving only a few effective degrees of freedom may erroneously find stable time-crystalline behaviour in the absence of such a noise suppression mechanism. Our proposal is that the role provided by long-range interactions can, however, be replaced by the effectively macroscopic rigidity of the ordered component of a symmetry-broken system such as the Ising magnet subjected to the dissipative processes discussed here.

Finally, we have only considered Markovian dissipation. An open and interesting problem is to understand whether the physics unveiled here is changed qualitatively in the non-Markovian case, and whether there are non-fine-tuned non-markovian environments that lead to interesting new examples of oscillatory dynamics in quantum systems.

\bibliography{refs}

\newpage~
\newpage

\onecolumngrid
\begin{center}
\textbf{Supplementary material: On time crystallinity in dissipative Floquet systems}\\
Achilleas Lazarides, Sthitadhi Roy, Francesco Piazza, and Roderich Moessner
\end{center}
\bigskip

\twocolumngrid

\setcounter{equation}{0}
\renewcommand{\theequation}{S\arabic{equation}}
\setcounter{figure}{0}
\renewcommand{\thefigure}{S\arabic{figure}}
\setcounter{page}{1}
\renewcommand{\thepage}{S\arabic{page}}

\section{Direct jump operators}

In this section of the supplementary material, we provide details of the direct jump operators [Eq.~\eqref{eq:dumpdef} of the main text] and describe how translation-invariance of the Hamiltonian as well as that of the jump operators can be exploited to simulate relatively large system sizes.
Note that the action of the direct jump operators, of the form 
$L_\alpha = \sqrt{\gamma}\left[\Theta(m_\alpha)\ket{\Up}\bra{\alpha} + \Theta(-m_\alpha)\ket{\Dn}\bra{\alpha}\right],$
on a basis state $\ket{\alpha}$ depends only on the magnetisation density of the state.
Similarly, the action of the $\hat{H}_\mathrm{TFIM}$ (in the $g\to0$ limit) as well as the global rotation operator $U_R(\theta)=\exp\left[-i\theta\sum_\ell\sigma^x_\ell\right]$ on $\ket{\alpha}$ also depends only on $m_\alpha$.
This allows for the fact that the density matrix elements in the Ising configuration basis depend only the magnetisation densities of the basis states. Mathematically, 
\begin{equation}
\rho_{\alpha\beta} = \varrho_{mm^\prime}~\forall (\alpha,\beta)~\mathrm{s.t.}~m_\alpha=m~\mathrm{and}~m_\beta=m^\prime.
\end{equation}
Hence, instead of the $2^N$-dimensional density matrix $\rho$, it suffices for us to work with the $(2N+1)$-dimensional matrix $\varrho$. Normalisation of the density matrix in this notation is ensured via $\sum_{m}\varrho_{mm}\mathcal{N}_m=1$, where $\mathcal{N}_m = \binom{N}{N(1+m)/2}$ is the multiplicity of the product states with magnetisation density $m$.
The probability distribution of the magnetisation density, $\phi_m$, is given by $\varrho_{mm}\mathcal{N}_m$ and the expectation values of the magnetisation $m$ and the correlation function $C$ are then given by
\begin{equation}
\begin{split}
\langle m\rangle &= \sum_{m}m\varrho_{mm}\mathcal{N}_m,\\
C &= \frac{1}{2}\sum_{m}\left(m^2-\frac{1}{N}\right)\varrho_{mm}\mathcal{N}_m,
\end{split}
\end{equation}
respectively.

\begin{figure}
\includegraphics[width=\columnwidth]{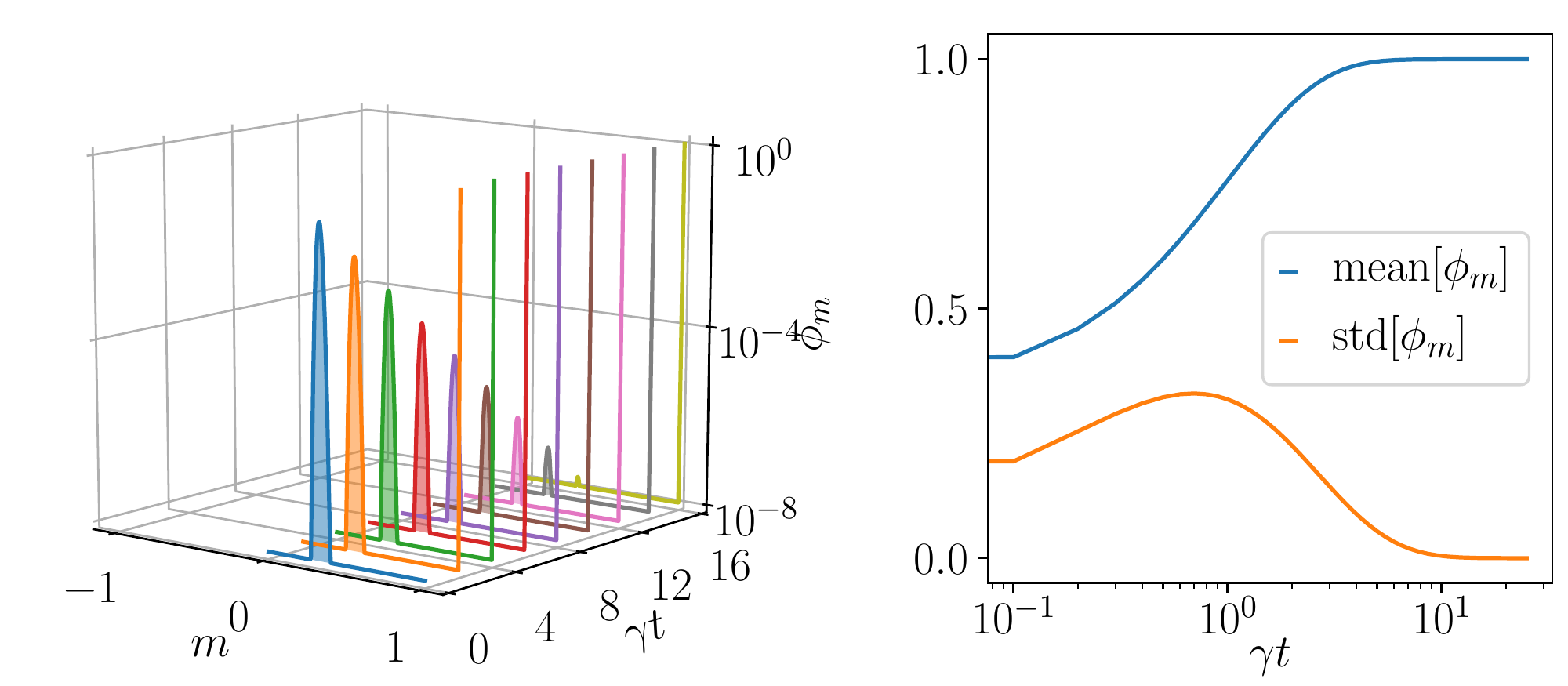}
\caption{{\bf Broadening of the state in magnetisation via the \emph{direct} jump operators:} The time-dependent distribution of the magnetisation as obtained from solving Eq.~\eqref{eq:dumpdiff} shows that the distribution is bimodal (and hence broad) for any finite cooling time. The right panel shows that the mean magnetisation density approaches unity as the system is rapidly cooled, however the standard deviation of the distribution $\phi_m$ is finite for all finite times.} 
\label{fig:broadeningdump}
\end{figure}

In order to solve for the dynamics of the system, we require the equation of the motion for $\varrho_{mm^\prime}$ under the action of the both, the unitary rotation operator as well as the jump operators.
The former is described by the equation
\begin{equation}
	\begin{split}
	\partial_t\varrho_{m,m^\prime} = -\frac{i}{2}N\theta[&(1-m^\prime)\varrho_{m,m^\prime+\frac{2}{N}}+(1+m^\prime)\varrho_{m,m^\prime-\frac{2}{N}}\\
	                                &-(1-m)\varrho_{m+\frac{2}{N},m^\prime}+(1+m)\varrho_{m-\frac{2}{N},m^\prime}].
	\end{split}
	\label{eq:unitarydiff}
\end{equation}
As that the jump operators transfer weight from all positive (negative) magnetisation states to the all-up (all-down) state directly with a rate $\gamma$, the evolution of $\varrho$ is described by the set of equations
\begin{equation}
\partial_t\varrho_{m,m^\prime} = \begin{cases}
                                  \gamma \sum\limits_{m^{\prime\prime}=1/N}^{1-1/N}\mathcal{N}_{m^{\prime\prime}}\varrho_{m^{\prime\prime},m^{\prime\prime}};\, m=m^\prime=1\\
                                  \gamma \sum\limits_{m^{\prime\prime}=-1/N}^{-1+1/N}\mathcal{N}_{m^{\prime\prime}}\varrho_{m^{\prime\prime},m^{\prime\prime}};\, m=m^\prime=-1\\
                                  -\gamma \varrho_{m,m^\prime};\, \mathrm{otherwise}
                                 \end{cases} ,
\label{eq:dumpdiff}
\end{equation}
To see that the jump operators spread the state out in magnetisation, one can simply analyse the solutions of Eq.~\eqref{eq:dumpdiff} with the initial conditions such that the state is narrowly distributed in magnetisation around a value $m(0)$. The solutions yield that all the $\varrho_{m,m^\prime}$ decay exponentially with a rate $\gamma$ except for $\varrho_{1,1}$ or $\varrho_{-1,-1}$ depending on if $m(0)\gtrless 0$, which approaches unity exponentially.
Specifically, for the initial condition $\varrho_{m(0)m(0)}(0)=1/\mathcal{N}_{m(0)}$, the solution to the time dependent distribution $\phi_m$ can be expressed as
\begin{equation}
\phi_m(t) = \delta_{m,m(0)}e^{-\gamma t} + \delta_{m,\pm1} (1-e^{-\gamma t}),
\end{equation}
for $m(0)\gtrless 0$, which manifestly shows that the time-dependent state is not sharply distributed in $m$ and the distribution has a finite standard deviation at any finite time $t$, see Fig.~\ref{fig:broadeningdump} for results.
Eqs.~\eqref{eq:dumpdiff} and \eqref{eq:unitarydiff} together describe the evolution of the density matrix of the system in the rotation cycle, whereas the former suffices in the cooling cycle if  we work in the $g\to0$ limit

\section{Domain-wall annihilating jump operators}
In this section, we discuss further details of the domain-wall annihilating operators. The operators in Eq.~\eqref{eq:localops} (main text) locally shift domain walls such that two of them can annihilate each other upon meeting. However the set of operators are rather inconvenient from the point of view of a numerical solution of the corresponding Lindblad equation as there are exponentially (in $N$) many jump operators. We surmount this problem by considering a different set of jump operators which have the same effect on basis product states as the ones in Eq.~\eqref{eq:localops} (main text) but which contains only polynomially in $N$ of them.
In particular, this new set contains $2^Z\times N$ jump operators, ($2^Z$ for each spin) where $Z$ is the coordination number of the lattice and $N$ is the number of spins.
As mentioned in the main text, each operator in the set consists of a spin-flip (or lowering or raising) term for a given site, say $\ell$, multiplied to projectors onto each configuration of its neighbours (hence $2^Z$ of them). If a neighbour-configuration has a net magnetisation which is positive (negative), the corresponding projector is multiplied to $\sigma^+_\ell$ ($\sigma^-_\ell$) and if the neighbour-configuration has equal number of up and down spins, the projector is multiplied to $\sigma^x_\ell$, the latter with much smaller rate.

For a square lattice in 2-$d$, for a given site $\ell$ at position $(x,y)$, there are three classes of jump operators:
\begin{itemize}
	\item All four of the neighbours are aligned with each other but anti-aligned with the spin at $\ell$. There are two jump operators for this case, $\mathcal{P}_{x+1,y+1}^\uparrow \mathcal{P}_{x+1,y-1}^\uparrow \mathcal{P}_{x-1,y+1}^\uparrow \mathcal{P}_{x-1,y-1}^\uparrow\sigma^+_{x,y}$ and the same with $\uparrow\rightarrow\downarrow$ and $+\rightarrow -$. See Fig.~\ref{fig:localops-2d_lattice} (right) [main text] for a visual example.
	\item Three of the neighbours are anti-aligned with the spin at $\ell$ and one is aligned. There are four configurations each for a the net magnetisation of the neighbours being positive or negative. So there are a total of eight operators in this class. As example, one of the operators in this class is $\mathcal{P}_{x+1,y+1}^\uparrow \mathcal{P}_{x+1,y-1}^\uparrow \mathcal{P}_{x-1,y+1}^\uparrow \mathcal{P}_{x-1,y-1}^\downarrow\sigma^+_{x,y}$, and the projector on the down spin could be on any of the four neighbours which generates the other three operators. Similarly $\uparrow\rightarrow\downarrow$ and $+\rightarrow -$ generates the other four operators in this class.
	\item Two of the neighbours are up and two of them are down. In this case, we always flip the spin at site $\ell$, but the operator has a much smaller rate. An example operator in this class is $\mathcal{P}_{x+1,y+1}^\uparrow \mathcal{P}_{x+1,y-1}^\uparrow \mathcal{P}_{x-1,y+1}^\downarrow \mathcal{P}_{x-1,y-1}^\downarrow\sigma^x_{x,y}$ There are six operators in this class (one for each of the two-up two-down configurations).
\end{itemize}
Note that the first two classes of the jump operators tries to reduce the length of the domain wall and favour the majority phase. However, they can get stuck if they encounter a straight domain wall. The third class of the jump operators serve to unfreeze such potentially frozen domain walls.

The effect of these jump operators on the diagonal elements of the density matrix can be understood from a classical Monte Carlo simulation using Glauber dynamics with the local energy cost function for a spin at site $\ell$ given by $\sigma^z_\ell\sum_{r\in\{r_\ell\}}\sigma^z_r$. While the Monte Carlo at zero temperature would try to generically take the system to the majority phase, there is a technical subtlety. The zero temperature classical Monte Carlo which tries to reduce the string tension of the domain wall can get stuck if it encounters a domain wall which straight; hence we run the Monte Carlo at a finite but small temperature. This essentially is the manifestation of the third class of the jump operators discussed above.

\begin{figure}
\includegraphics[width=\columnwidth]{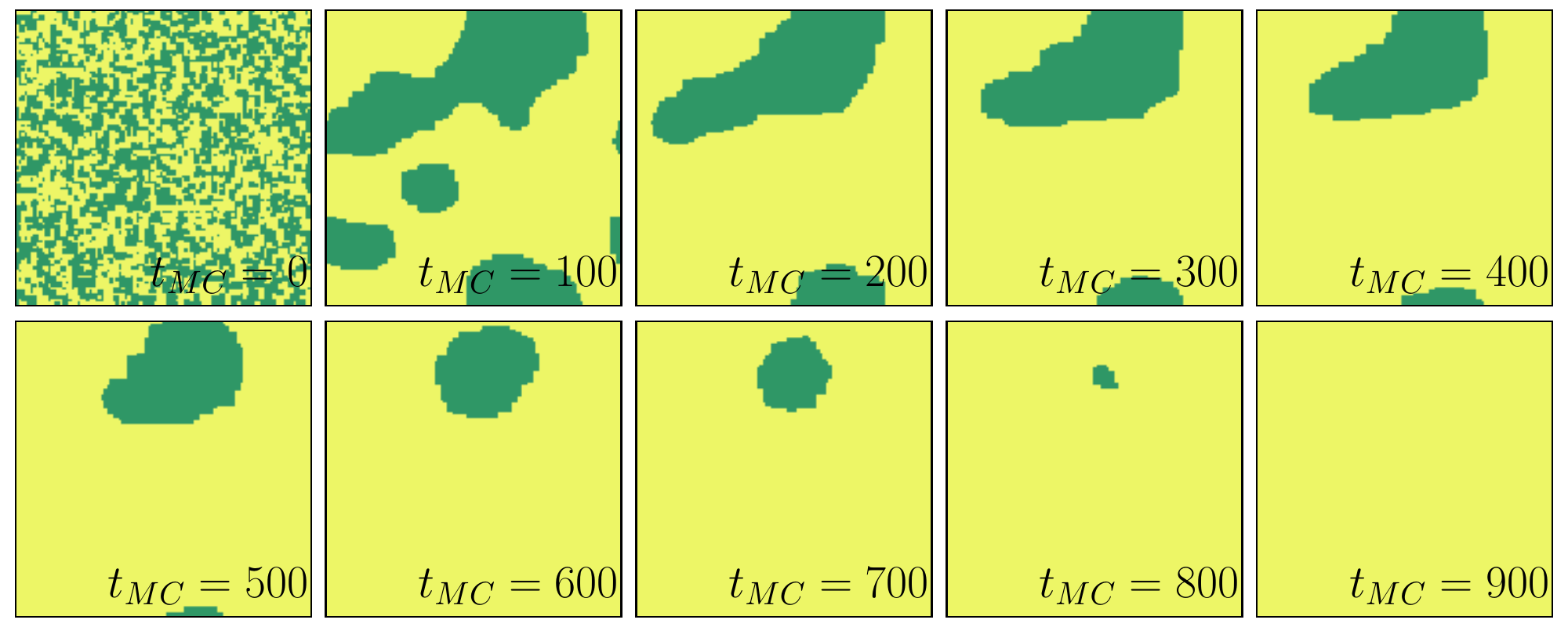}
\caption{{\bf Shrinking of domain walls in classical Monte Carlo:} Evolution of the spin configuration  on a 100$\times$100 lattice obtained using classical Monte Carlo dynamics (Glauber dynamics) at inverse temperature, $\beta=2$ visually shows the shrinking of the domain walls and eventual approach to a configuration completely taken over by the majority phase. Green regions denote spin-down and yellow regions, spin-up.}
\label{fig:coarsening}
\end{figure}

Fig.~\ref{fig:coarsening} shows a specific Monte Carlo trajectory with each panel showing the spin-configuration on the lattice at specific Monte Carlo times; green denotes regions of down-spin and yellow, up-spin. The evolution of the configuration clearly shows the shrinking of the domain-wall lengths and corresponds to a macroscopic coarse-grained version of the processes shown in Fig.~\ref{fig:localops-2d_lattice} (main text).

As a final remark, we mention that the full solution of the Lindblad equation in the 2-$d$ case (results of Fig.~\ref{fig:local2d} (main text)) was obtained using the Monte Carlo wave-function method, see [M{\o}lmer {\it et al.}, J. Opt. Soc. Am. B {\bf{10},} 524 (1993)] for details.

\end{document}